\documentclass[usenatbib]{emulateapj}

\usepackage{graphicx}
\usepackage{epsfig,psfrag}
\usepackage{natbib}
\usepackage{multirow}

\def\hinvMpc{\;h^{-1}\;{\rm {Mpc}}}

\shorttitle{THE HOD OF SDSS QUASARS}
\shortauthors{Richardson et al.}

\begin{document}
\title{The Halo Occupation Distribution of SDSS Quasars} 
\author{Jonathan Richardson$^{1,3}$, Zheng Zheng$^{2,3}$, Suchetana Chatterjee$^{1,4}$, Daisuke Nagai$^{1,3}$, Yue Shen$^{5}$}
\affiliation{$^{1}${Department of Astronomy, Yale University, New Haven, CT 06520.}\\
             $^{2}${Department of Physics and Astronomy, University of Utah, Salt Lake City, UT 84112.}\\
             $^{3}${Department of Physics, Yale University, New Haven, CT 06520.}\\
             $^{4}${Department of Physics and Astronomy, University of Pittsburgh, Pittsburgh, PA 15260.}\\
  	     $^{5}${Harvard-Smithsonian Center for Astrophysics, 60 Garden Street, MS-51, Cambridge, MA 02138.}\\
	    }

\begin{abstract}
We present an estimate of the projected two-point correlation function (2PCF) of quasars in the Sloan Digital Sky Survey (SDSS) over the full range of one- and two-halo scales, $0.02 \; h^{-1} \; \mathrm{Mpc} < r_p < 120 \; h^{-1} \; \mathrm{Mpc}$. This was achieved by combining data from SDSS DR7 on large scales and \citet[][with appropriate statistical corrections]{HE06} on small scales. Our combined clustering sample is the largest spectroscopic quasar clustering sample to date, containing $\sim 48,000$ quasars in the redshift range $0.4 \lesssim z \lesssim 2.5$ with median redshift 1.4. We interpret these precise 2PCF measurements within the halo occupation distribution (HOD) framework and constrain the occupation functions of central and satellite quasars in dark matter halos. In order to explain the small-scale clustering, the HOD modeling requires that a small fraction of $z\sim 1.4$ quasars, $f_{\mathrm{sat}}=(7.4 \pm 1.4) \times 10^{-4}$, be satellites in dark matter halos. At $z\sim 1.4$, the median masses of the host halos of central and satellite quasars are constrained to be $M_{\mathrm{cen}} = 4.1^{+0.3}_{-0.4} \times 10^{12} \; h^{-1} \; \mathrm{M_{\sun}}$ and $M_{\mathrm{sat}} = 3.6^{+0.8}_{-1.0}\times 10^{14} \; h^{-1} \; \mathrm{M_{\sun}}$, respectively. To investigate the redshift evolution of the quasar-halo relationship, we also perform HOD modeling of the projected 2PCF measured by \citet{SH07} for SDSS quasars with median redshift 3.2. We find tentative evidence for an increase in the mass scale of quasar host halos---the inferred median mass of halos hosting central quasars at $z\sim3.2$ is $M_{\mathrm{cen}} = 14.1^{+5.8}_{-6.9} \times 10^{12} \; h^{-1} \; \mathrm{M_{\sun}}$. The cutoff profiles of the mean occupation functions of central quasars reveal that quasar luminosity is more tightly correlated with halo mass at higher redshifts. The average quasar duty cycle around the median host halo mass is inferred to be $f_{\mathrm{q}}=7.3^{+0.6}_{-1.5}\times 10^{-4}$ at $z\sim1.4$ and $f_{\mathrm{q}}=8.6^{+20.4}_{-7.2}\times 10^{-2}$ at $z\sim3.2$. We discuss the implications of our results for quasar evolution and quasar-galaxy co-evolution.
\end{abstract}

\section{Introduction}

\begin{table*}[t]
\begin{center}
\caption{SDSS Quasar Clustering Samples}
\setlength{\tabcolsep}{8pt}
\begin{tabular}{l c c c c c c c}
\hline
\hline\noalign{\smallskip}
Sample & $z$ & $\bar{z}$ & Flux Limit & Area [$\mathrm{deg}^{2}$] & $N_{\mathrm{QSO}}$ & $\pi_{\mathrm{max}}$ [$h^{-1}$ Mpc] & $r_{p}$ [$h^{-1}$ Mpc]\\
\noalign{\smallskip}\hline\noalign{\smallskip}
SDSS DR7 & $0.4 < z < 2.5$ & 1.4 & $i < 19.1$ & 6248 & 47,699 & 80.0 & $1.0 < r_{p} < 120$\\[0.05cm]
Shen et al. (2007) & $2.9 < z < 5.4$ & 3.2 & $i < 20.2$ & 4041 & 4426 & 100.0 & $1.3 < r_{p} < 211$\\[0.05cm]
Hennawi et al. (2006) & $0.7 < z < 3.0$ & 1.6 & $i < 19.1/21.0$ & \nodata & 386 & $\sim22$ & $0.02 < r_{p} < 7.0$\\
\noalign{\smallskip}\hline
\end{tabular}
\end{center}
\label{tab:tab1}
\tablecomments{
For each sample, the $z$ column indicates the redshift range, $\bar{z}$ is the median redshift, $N_{\mathrm{QSO}}$ is the total number of quasars, $\pi_{\mathrm{max}}$ is the upper limit of the line-of-sight pair separation for computing the projected 2PCF [see eq.~(\ref{eqn:1})], and the $r_{p}$ column is the (comoving) transverse pair separation range.
}
\end{table*}

Quasars are a highly luminous class of active galactic nuclei (AGN) believed to be powered by supermassive black holes \citep[e.g.,][]{SA64, LY69, KO95, RI98}. Theoretical studies and observational evidence indicate that an epoch of quasar activity occurred during the formation of every massive elliptical galaxy \citep[e.g.,][]{S&M96, HO08}. Correlations between black hole mass and host galaxy properties \citep[e.g.,][]{MA98,  FE00, GE00, ME01, GR02, TR02, KI03} suggest that black holes and galaxy spheroids evolve via a common physical process \citep[e.g.,][]{HO08}. More generally, it has been suggested that there exists a black hole fundamental plane \citep{hopkinsetal07} analogous to the fundamental plane for elliptical galaxies \citep{dressleretal87, D&D87}, although recent results provide only weak evidence for its existence \citep{beifiori12}. Combining these ideas, \citet{HO08} showed using a hydrodynamic simulation that in a major merger-driven scenario, galaxies evolve from an ultraluminous infrared galaxy to an elliptical galaxy through a quasar phase. In this scenario quasars play an important role in the observed bi-modality of galaxy color, as strong radiative feedback from quasars blows out gas from galaxies and suppresses star formation \citep[e.g.,][]{SI98, KA00, VO03, WY03, AD05, DI05, CR06, HO06a}.

Galaxies are known to reside in dark matter halos and, as such, probe structure formation in the universe \citep[e.g.,][]{w&f91, kauffmannetal93, nfw95, m&w96, kauffmannetal99, springeletal05}. Since supermassive black holes are believed to reside at the centers of massive galaxies \citep[e.g.,][]{SO82}, a connection between black holes and their host dark matter halos is naturally expected. It has recently been suggested that this connection may be indirect, however, as supermassive black hole masses have been found to strongly correlate only with bulge properties \citep{kormendy11}. The black hole-halo relationship has been studied in analytic models and cosmological hydrodynamic simulations \citep{b&s09, b&s10, dimatteoetal08, volonterietal11}. Observationally, quasars provide a powerful tool for studying this relation.

Because of their high luminosity, quasars are detected to $z \gtrsim 7$ \citep[e.g.,][]{mortlocketal11}, making them excellent probes of structure formation over cosmic time. The spatial clustering of quasars can be used to probe the relationship between quasars and their host dark matter halos, providing constraints on the formation and evolution of quasars and their role in galaxy formation. In this paper, we present clustering measurements of a large spectroscopic sample of quasars ($\sim 48,000$ quasars in the redshift range $0.4 \lesssim z \lesssim 2.5$) from the Sloan Digital Sky Survey \citep[SDSS;][]{YO00}, spanning scales $\sim 0.02\hinvMpc$ to $\sim 120 \hinvMpc$, and perform theoretical modeling of the clustering to infer the relation between quasars and dark matter halos.

Quasar clustering is typically measured through the two-point correlation function \citep[2PCF; e.g.,][]{AR70}. The advent of the 2dF Quasi-Stellar Object Redshift Survey \citep[2QZ;][]{CR04a} and the SDSS has enabled high precision measurements of the 2PCF \citep[e.g.,][]{PO04, CR05, DA05, MY06, MY07a, MY07b, PO06, SH07, SH08, SH09, DA08, RO09, hickox11, white12}. It was found that quasars become increasingly biased relative to the underlying dark matter with increasing redshift \citep[e.g.,][]{CR01,PO04,CR05,MY07a,SH07}, and that quasar clustering only weakly depends on luminosity \citep[e.g.,][]{CR05,PO06,MY07a,DA08,SH09}.

With an assumed cosmological model, the relative amplitude of the large-scale quasar 2PCF and dark matter 2PCF (i.e., the square of the quasar bias) can be inferred. By relating this bias factor to that of dark matter halos \citep[e.g.,][]{jing98,shethetal01}, a rough estimate of the typical mass of a quasar-hosting dark matter halo can then be obtained. Observed quasar clustering shows that the typical halo mass is in the range of $10^{12}-10^{13} \; h^{-1} \; \mathrm{M}_{\sun}$, independent of redshift and luminosity \citep[e.g.,][]{PO04,CR05,MY06,MY07a,coiletal07, SH07,DA08,PA09, hickoxetal09, white12}. Theoretically, the characteristic mass of halos hosting quasars may be related to the critical mass beyond which halos do not have large reservoirs of cold gas \citep[e.g.,][]{HO06a, crotonetal09, keresetal09}. From the quasar abundance, the abundance of halos within the host halo mass range, and the halo formation time, the quasar lifetime (and duty cycle) can also be constrained \citep{cole89,MA01, HA01, Shankar09, Shankar10a, Shankar10b, Shankar11}.

Measurement of the small-scale quasar clustering is usually hindered by fiber collisions in fiber-based spectroscopic surveys. On a single spectroscopic plate in the SDSS, the finite size of the fiber plugs prevents any two fibers from being placed within $\sim 55 ''$ of each other, corresponding to a comoving separation of $\sim 1 \hinvMpc$ at the typical quasar redshift of $1.5$. \citet[]{MY06} circumvented this problem by measuring the angular clustering of photometrically classified quasars on all scales, and \citet[][hereafter HE06]{HE06} obtained the first measurement of the real-space quasar clustering on small scales by detecting close quasar binaries through follow-up spectroscopy of SDSS quasars. HE06 detected excess clustering at small scales over extrapolations of the power law from large scales. This result was confirmed by \citet{myers08}, who found a slightly smaller excess than HE06 using a more homogeneously selected sample. At small scales, quasar clustering probes the distribution of quasars within dark matter halos. It can advance our understanding of the triggering processes that drive quasar accretion \citep[e.g.,][]{HO08}. The distribution of quasars inside halos inferred from small-scale clustering can also provide insight into black hole mergers and binary AGN \citep[e.g.,][]{comerfordetal09, liuetal10, fuetal11, liuetal11, mcgurketal11}.

Accurate inferences of physical parameters, such as the host halo mass, satellite fraction, and quasar lifetime, rely on precision clustering measurements. Specific predictions from theory \citep[e.g., redshift and luminosity evolution of quasar lifetimes:][]{cole89,MA01,shen09,crotonetal09, bonolietal09} require precision measurements for informative tests. At present, large uncertainties in the host halo mass scale prevent firm establishment of the ``no redshift evolution'' scenario \citep[e.g.,][]{HO08, hickoxetal09}. Clustering of X-ray AGN suggests that the host halos of X-ray AGN are more massive than the hosts of quasars \citep[e.g.,][]{gillietal05, coiletal09}, but the constrained mass range is too large for a definite conclusion \citep[for a recent review, see][]{cappelluti12}.

We now present precise quasar 2PCF measurements covering small to large scales and model the halo occupation distribution (HOD) to obtain accurate constraints on the host halo mass scale of quasars and its redshift evolution. We also obtain the quasar satellite fraction and the quasar duty cycle from our HOD analysis. The HOD provides a powerful theoretical framework \citep[e.g.,][]{BE02} for understanding the clustering properties of any biased tracer of mass. It has been used extensively for modeling and understanding galaxy clustering \citep[e.g.,][]{SE00, ZH05, ZH07, Zh09} and extended to study AGN and quasar clustering \citep[e.g.,][]{PO04, wake08, MI11, SHE10, ST11}. The work presented in this paper adopts, for the first time, a theoretically motivated model of the AGN HOD to interpret the spatial clustering of quasars over a large range of scales.

Our paper is organized as follows. In $\S 2$, we briefly describe our data sets. In $\S 3$, we present our estimate for the full projected 2PCF of SDSS quasars. The parameterization of the quasar HOD and the theoretical model of the 2PCF are presented in $\S 4$. The HOD modeling of the 2PCF and results are presented in $\S 5$. Finally, we discuss our results in $\S 6$ and provide a summary in $\S 7$. Throughout the paper we assume a spatially flat, $\Lambda$CDM cosmology: $\Omega_{m}=0.26$, $\Omega_{\Lambda}=0.74$, $\Omega_{b}=0.0435$, $n_{s}=0.96$, $\sigma_{8}=0.78$, and $h=0.7$ \citep{SP07}. We quote all distances in comoving $h^{-1}$ Mpc and masses in units of $h^{-1}$ $\mathrm{M}_{\sun}$. All magnitudes are in the $\mathrm{AB_{95}}$ system \citep{FU96}.

\section{Data}
The projected 2PCF measurements are drawn from three quasar clustering samples, a low-redshift sample (median redshift 1.4) from SDSS DR7,  a high-redshift sample (median redshift 3.2) from SDSS DR5, and a binary quasar sample (median redshift 1.6) for small-scale clustering. The samples are summarized in Table~\ref{tab:tab1} and described below.

\subsection{SDSS Quasars}
The DR7 spectroscopic quasar catalog \citep{SC10,Shen11} serves as the parent data set from which we build our clustering sample. This catalog contains 105,783 spectroscopically confirmed quasars with $M_{i} < -22.0$ over an area of $9380 \; \mathrm{deg^2}$ and a redshift range of $0.065<z<5.46$. In order to construct a homogeneous clustering sample, we only consider quasars over the $6248 \; \mathrm{deg^2}$ covered by the final target selection algorithm \citep[e.g.,][]{RI02}. Quasars at $z < 3.0$ are flux limited to $i < 19.1$, while higher-redshift quasars, which are fainter and rarer, are flux limited to $i < 20.2$. Additionally, we restrict our clustering sample to quasars in the redshift range $0.4<z<2.5$. Our clustering sample then consists of 47,699 quasars flux limited to $i < 19.1$, with a median redshift of $\bar{z}=1.4$.

At high redshifts, \citet[][hereafter SH07]{SH07} have measured the projected 2PCF of quasars in SDSS DR5 \citep{AD07, SC07}. The sample consists of 4426 spectroscopically identified quasars in the redshift interval $2.9 < z < 5.4$, with a median redshift of $ \bar{z} = 3.2$. This sample is uniformly selected using the same targeting algorithm as \citet{RI02}. SH07 calculate the projected 2PCF using the estimator of \citet{LAN93} and adopting a maximum line-of-sight separation of $\pi_{\rm max}=100.0$ $h^{-1}$ Mpc. They measure the 2PCF over the transverse separation range 1.3 $h^{-1}$ Mpc $<r_{p} < 211$ $h^{-1}$ Mpc. We refer the reader to SH07 for a full discussion.

\subsection{Binary Quasars}
HE06 present measurements of the projected 2PCF of binary quasars to separations as small as $r_{p} \sim 20$ $h^{-1}$ kpc. The clustering sample spans the redshift range $0.7 \le z \le 3.0$ with a median redshift of $ \bar{z} \approx 1.6$. In studying binary quasars, HE06 include in their clustering sample only same-redshift quasar pairs, defined as pairs with velocity differences of $ \mid \Delta v \mid < 2000$ km $\mathrm{s}^{-1}$. HE06 detect faint companions around a parent sample of 52,279 SDSS DR3 quasars and 6879 2QZ quasars with matching SDSS photometry. They also identify binary quasars from a photometric sample of 273,287 SDSS DR3 quasar candidates, but we do not discuss them here because these pairs were not included in the clustering sample. In order to construct a sample of close quasar pairs large enough for a clustering study, HE06 employ four target selection algorithms, each with a different flux limit and varying degree of completeness, applied over different angular separation scales. We briefly discuss these algorithms below. We refer the reader to HE06 for a full discussion.

For angular separations of $\theta \le 3''$, pairs are identified by fitting a multi-component point-spread function to atlas images of each quasar in a subset of 39,142 SDSS quasars \citep[e.g.,][]{PI03, IN03}. For angular separations $3'' < \theta \le 60''$, HE06 employ a color similarity statistic exploiting the quasar color-redshift relation \citep[e.g.,][]{RI01}. In both algorithms, companion objects are flux limited to $i < 21.0$ for the follow-up spectroscopy required to confirm the candidate quasar pair. At separations of $\theta > 60''$, HE06 restrict their search to the 52,279 quasars in the SDSS spectroscopic catalog, where an observational completeness of $\sim 95 \%$ has been measured for $i < 19.1$ \citep{VA05}. No additional selection criteria are imposed beyond those of the SDSS primary target selection algorithm.

\section{The Full Projected Two-Point Correlation Function}

\begin{figure}
\begin{center}
\begin{tabular}{c}.
\includegraphics[width=8cm]{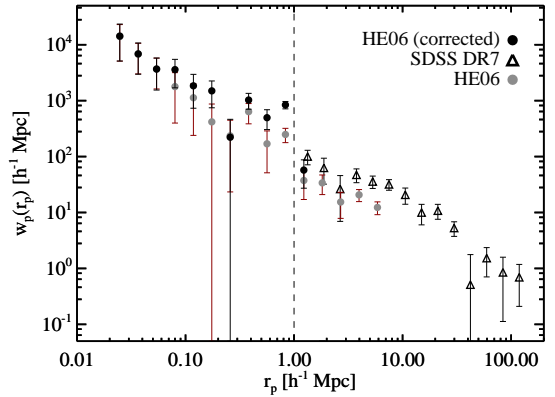} 
\end{tabular}
\caption{
\label{fig:fig1}
Projected 2PCF of the low-redshift quasar sample (median redshift 1.4). Triangles represent our measurements with the SDSS DR7 data. The original estimates from HE06 with the binary quasar sample are shown as light filled circles. Our correction to the HE06 estimates by adopting $\pi_{\mathrm{max}}=80\hinvMpc$, as used in our SDSS DR7 measurements, leads to the dark filled circles. The dashed vertical line indicates the fiber collision scale for the median redshift.
}
\end{center}
\end{figure}

For the high-redshift sample, we adopt the measurements in SH07. For the low-redshift sample, however, we combine two sets of measurements to form the full projected 2PCF of quasars at $\bar{z}=1.4$. We measure the projected 2PCF of quasars in SDSS DR7 on scales above $1\hinvMpc$. To accurately measure the projected 2PCF on scales below $1\hinvMpc$, we make use of the binary quasar sample in HE06, with necessary corrections.

To measure the large-scale 2PCF of SDSS DR7 quasars, we follow the standard practice \citep[e.g.,][]{ZE05,SH07} to generate random catalogs using the angular selection function and redshift distribution of the DR7 uniform quasar sample. We compute the redshift-space 2PCF $\xi_{s}(r_{p},\,\pi)$ using the Landy-Szalay estimator \citep{LAN93}, with $r_p$ and $\pi$ the transverse and line-of-sight pair separations, respectively. The projected 2PCF $w_p$ is obtained by integrating along the line of sight \citep{DA83},
\begin{equation}
\label{eqn:1}
w_{p}(r_{p})=2 \int_{0}^{\pi_{\mathrm{max}}} \xi_{s}(r_{p},\,\pi) d\pi,
\end{equation}
where we adopt an upper limit of $\pi_{\mathrm{max}}=80 \hinvMpc$ for the line-of-sight separation. To estimate errors on the 2PCF, we use jackknife resampling as described in \citet{SH07} with twenty-five jackknife samples. The measurements are shown as triangles with error bars in Figure~\ref{fig:fig1}. A more detailed analysis of quasar clustering based on SDSS DR7 quasars as functions of redshift and physical properties will be presented in a forthcoming paper (Shen et al., in preparation).

The 2PCF measurements obtained from the DR7 sample become unreliable at pair separations $\lesssim 1$ $h^{-1}$ Mpc because of fiber collisions. We overcome this limitation using data from HE06 for small scales, enabling us to measure the 2PCF down to transverse separations of $\sim 0.02\hinvMpc$. We correct for the dominant selection difference between our samples as outlined below and in the Appendix.

On average, quasars in the HE06 sample are one magnitude fainter and at slightly higher redshifts compared to those in our SDSS DR7 sample. There is also a difference in the maximum line-of-sight separation, $\pi_{\mathrm{max}}$, used in calculating the projected 2PCF. In the original HE06 estimates, $\pi_{\mathrm{max}}$ corresponds to $\sim 22 \hinvMpc$, while our DR7 measurements adopt $\pi_{\mathrm{max}}=80\hinvMpc$. Figure~\ref{fig:fig1} shows the combined effect of these sample selection differences on the projected 2PCF measurements. Over transverse separations 1-7 $h^{-1}$ Mpc, the scales over which the two sets of measurements overlap, our SDSS DR7 measurements (triangles) exceed the HE06 measurements (light filled circles) by an average factor of 2.3. The bulk of this discrepancy can be attributed to the lower value of $\pi_{\mathrm{max}}$ used in HE06, which causes the offset to increase with $r_p$. This effect is less significant on small scales ($r_{p} \ll 1$ $h^{-1}$ Mpc) because the contribution to the projected 2PCF from pairs with large line-of-sight separations is small, allowing the clustering signal to be well recovered even with a smaller $\pi_{\mathrm{max}}$.

In the Appendix, we discuss in detail our correction for the difference in $\pi_{\mathrm{max}}$. The corrected HE06 projected 2PCF measurements, indicated by the dark filled circles in Figure~\ref{fig:fig1}, are consistent with the SDSS DR7 data to within $1\sigma$. HE06 only catalog projected quasar pairs (on which our correction relies) with angular separations $<90''$, so our correction cannot be applied to the data points above $\sim 1.5 \hinvMpc$. Combining the corrected HE06 measurements for $r_{p} < 1 \; h^{-1} \; \mathrm{Mpc}$ with our SDSS DR7 measurements for $r_{p} \ge 1.0$ $h^{-1}$ Mpc, we obtain an estimate of the full projected 2PCF of SDSS quasars, which is presented in Table~2. The uncertainties on each measurement are listed under $\delta w_{p}$. For comparison, we also list the original measurements of HE06 in parentheses. We discuss the robustness of this small-scale 2PCF estimate in $\S 6.1$ and its possible luminosity evolution in \S6.2.

\begin{table}[t]
\begin{center}
\caption{The Full Projected 2PCF of Low-$z$ ($\bar{z}=1.4$) SDSS Quasars}
\setlength{\tabcolsep}{8pt}
\begin{tabular}{r r r r r}
\hline
\hline\noalign{\smallskip}
$r_p$ & $w_p$& $w_p$[HE06] & $\delta w_{p}$ & $\delta w_{p}$[HE06]\\
\noalign{\smallskip}\hline\noalign{\smallskip}
  0.025 & 14270  & (14020) & 9158 & (8928)\\[0.05cm]
  0.037 &  6864   & (6803) & 3892 & (3795)\\[0.05cm]
  0.054 &  3668   & (3688) & 2122 & (2068)\\[0.05cm]
  0.080 &  3588   & (1783) & 1894 & (1386)\\[0.05cm]
  0.118 &  1842   & (1128) & 1110 & (888.4)\\[0.05cm]
  0.175 &  1497   &(417.6) & 751.5 & (453.4)\\[0.05cm]
  0.258 &   221.6 & (235.0) & 242.2 & (211.6)\\[0.05cm]
  0.381 &  1026  & (636.2) & 323.8 & (251.0)\\[0.05cm]
  0.564 &   496.0 & (169.4) & 192.2 & (118.1)\\[0.05cm]
  0.833 &   839.8 & (248.6) & 126.8 & (71.08)\\[0.05cm]
  1.334 &   100.3 & & 29.16 & \\[0.05cm]
  1.884 &   62.82 & & 30.66 & \\[0.05cm]
  2.661 &   26.35 & & 19.41 & \\[0.05cm]
  3.758 &   47.19 & & 13.06 & \\[0.05cm]
  5.309 &   35.87 & & 8.942 & \\[0.05cm]
  7.499 &   31.83 & & 6.844 & \\[0.05cm]
 10.593 &   20.73 & & 6.668 & \\[0.05cm]
 14.962 &   10.00 & & 3.988 & \\[0.05cm]
 21.135 &   10.78 & & 3.216 & \\[0.05cm]
 29.854 &    5.283 & & 1.543 & \\[0.05cm]
 42.170 &    0.5126 & & 1.255 & \\[0.05cm]
 59.566 &    1.531 & & 0.8258 & \\[0.05cm]
 84.140 &    0.8502 & & 0.7377 & \\[0.05cm]
118.850 &    0.6947 & & 0.4851 & \\
\noalign{\smallskip}\hline
\end{tabular}
\end{center}
\tablecomments{
The values of $r_p$, $w_p$, and $\delta w_p$ (error bar) are all in units of $\hinvMpc$. The third and fifth columns are from the original measurements of HE06 below $1\hinvMpc$ (see the text).}
\end{table}

\section{Theoretical Framework for Modeling Quasar Clustering}
In this paper, we interpret the quasar 2PCF measurements within the HOD framework. The HOD provides a complete description of the relation between quasars and dark matter at the level of individual virialized halos. In the following sections, we introduce our quasar HOD parameterization and describe the calculation of the 2PCF from the HOD.

\subsection{Halo Occupation Distribution of Quasars}
In analogy to the galaxy HOD \citep[e.g.,][]{BE02}, the quasar HOD is defined by $P(N|M)$, the conditional probability that a halo of virial mass $M$ contains $N$ quasars above some specified luminosity threshold, and by the relative spatial and velocity distributions of quasars within halos. In principle, $P(N|M)$ could be fully specified by determining all its moments observationally from the quasar clustering at each order. In reality, however, quasar samples are too sparse to reliably measure higher-order clustering statistics. For our purpose of modeling the 2PCF, we only need the description of the first two moments, $\langle N(M) \rangle$ and $\langle N(N-1) \rangle_{M}$. Here we assume that the quasar HOD depends on halo mass alone, i.e., the quasar content of halos at a given mass is statistically independent of the large-scale environments within which those halos reside \citep[e.g.,][]{BO91, LK99}. We neglect the assembly bias effect \citep[e.g.,][]{GA05}, which should be small for the massive halos that typically host quasars.

We represent the quasar mean occupation function as the sum of its physically illustrative central and satellite components, $\langle N(M) \rangle = \langle N_{\mathrm{cen}}(M)\rangle + \langle N_{\mathrm{sat}}(M)\rangle$ \citep[e.g.,][]{KR04,ZH05,Chatterjee12}. Our parameterization is largely based on the model of \citet{Chatterjee12}, which results from a study of low-luminosity AGN in a cosmological hydrodynamic simulation \citep{DI08}. The mean occupation function is given as a softened step function for the central component plus a rolling-off power law for the satellite component,
\begin{eqnarray}
\label{eqn:2}
\langle N(M)\rangle &=& \frac{1}{2}\left[1+{\rm erf}\left(\frac{{\rm log} M-{\rm log} M_{\rm{min}}}{\sigma_{\rm{log M}}}\right)\right] \nonumber \\ & + & \left(\frac{M}{M_{1}}\right)^{\alpha} \exp \left(-\, \frac{M_{\mathrm{cut}}}{M} \right).
\end{eqnarray}
The model admits five free parameters: $M_{\mathrm{min}}$, the characteristic mass scale at which on average half of the halos host one quasar at the center of each halo; $\sigma_{\log M}$, the characteristic transition width of the softened step function; $M_{1}$, the approximate mass scale at which halos host, on average, one \textit{satellite} quasar; $\alpha$, the power law index of the mean satellite occupation function; and $M_{\mathrm{cut}}$, the mass scale below which the satellite mean occupation decays exponentially (see \S6.3 for a discussion of alternative HOD models). We assume that the halo occupations of central and satellite quasars are independent. That is, for a given halo, the occupation of satellite quasars does not depend on whether there is a central quasar in the halo. This assumption is motivated by the results in \citet{Chatterjee12}, who find no correlation between the activities of central and satellite AGN in a hydrodynamic simulation.
 
For a given halo mass, satellite galaxies and AGN in simulations are found to follow a Poisson distibution \citep[e.g,][]{ZH05, DE11, Chatterjee12}. Hence we adopt a Poisson distribution for the satellite occupation number and a nearest integer distribution \citep{BE02} for the central quasar occupation number. These are used to compute the second moment of the occupation number in the theoretical model of the 2PCF (see below). The above HOD parameterization is the fiducial model used in this paper, but in \S 6.3 we also explore several alternate parameterizations to test our results.

For the spatial distribution of satellite quasars within halos, we assume an NFW profile \citep{NA97} with the concentration-mass relation \citep{ZH07}
\begin{equation}
c(M,\,z)= \frac{c_{0}}{1+z} \left( \frac{M}{M_{*}} \right)^{\beta},
\end{equation}
where $M_{*}$ is the nonlinear mass for collapse at $z=0$, $c_{0}$ is the concentration parameter, and $\beta=-0.13$. We adopt $c_{0}=25$ for modeling $z\sim 1.4$ quasars. Assuming that the concentration increases with redshift, we adopt $c_{0}=45$ for modeling $z\sim 3.2$ quasars. These values are motivated by the high concentration seen in observations of AGN \citep[]{LI07}. We have verified that our modeling only weakly depends on $c_{0}$ for a wide range of values, from $\sim10$ to $\sim60$. We note that an NFW profile with high concentration is similar to a power law with index -3, which has been seen in the spatial distribution of black holes and low-luminosity AGN in hydrodynamic simulations \citep[]{DE11, Chatterjee12}.

\subsection{2PCF Calculation}
The quasar 2PCF, $\xi_{q}(r)$, gives the excess probability above random of finding a pair of quasars separated by a distance $r$. It can be represented in terms of the contributions from intra-halo pairs, $\xi_{1h}(r)$, and inter-halo pairs, $\xi_{2h}(r)$. The inter-halo or two-halo term is approximated as \citep[e.g.,][]{BE02}
\begin{equation}
\xi_{2h}(r) \approx \biggl[n_{q}^{-1} \int_{0}^{\infty} dM \frac{dn}{dM} \langle N(M)\rangle b_{h}(M)\biggr]^{2} \xi_{m}(r),
\end{equation}
where $n_{q}$ is the quasar number density, $dn/dM$ is the differential halo mass function, $b_{h}(M)$ is the halo bias factor, and $\xi_{m}(r)$ is the 2PCF of matter. We identify the bracketed term as the quasar linear bias factor, $b_{q}$. The intra-halo or one-halo term is expressed as
\begin{equation}
1+\xi_{1h}(r) \approx \frac{1}{4 \pi n_{q}^{2} r^{2}} \int_{0}^{\infty} dM \frac{dn}{dM} \left\langle N\left(N-1\right)\right\rangle _{M} \frac{dF_M}{dr},
\end{equation}
where, for halos of mass $M$, $F_M(r)$ is the average fraction of same-halo pairs at separations $\le r$. The two-halo term depends only on $\langle N(M) \rangle$, while the one-halo term depends on $\left\langle N\left(N-1\right)\right\rangle _{M}$ and the radial profile of the spatial distribution of quasars through $F_M(r)$.

In detail, the 2PCF calculation in our model is more complicated than the above expressions, as it includes the effects of halo exclusion, nonlinear clustering, and scale-dependent halo bias. It follows the method proposed in \citet{TI05}, which improves the algorithm in \citet{ZH04} by incorporating a more accurate treatment of the halo exclusion effect. Halos in our calculation are defined as objects with mean density of 200 times that of the background universe. The halo mass function is computed according to the formula given by \citet{Jenkins01}. For the large scale halo bias factor, we adopt the formula in \citet{TI05}.

\section{HOD Modeling and Analysis}
Using the routine developed in \citet{ZH07}, we perform a Markov Chain Monte Carlo (MCMC) modeling of the projected 2PCF to sample the five-dimensional parameter space of the quasar HOD. We adopt flat priors in logarithmic space for $M_{\mathrm{min}}$, $M_{1}$, and $M_{\mathrm{cut}}$ and in linear space for $\alpha$ and $\sigma_{\log M}$ ($\sigma_{\log M} > 0$). We additionally require $M_{\mathrm{cut}}>10^{12}$ $h^{-1}$ $\mathrm{M}_{\sun}$ and $0.5 < \alpha < 4.0$, as motivated by theoretical studies and computational efficiency. We fit both the 2PCF and the abundance (number density) of quasars to constrain the HOD. The quasar number density and its associated uncertainty are obtained from the luminosity function of \citet{SH12}, evaluated at the flux limit and median redshift of each sample. Since our clustering samples are sparse, we calculate $\chi^{2}$ values using only the diagonal elements of the covariance matrix. Each MCMC chain contains 100,000 points in the parameter space for modeling the low-redshift and high-redshift samples.

\subsection{Best-Fit HOD}

\begin{figure*}
\begin{center}
\begin{tabular}{c}
\includegraphics[width=16cm]{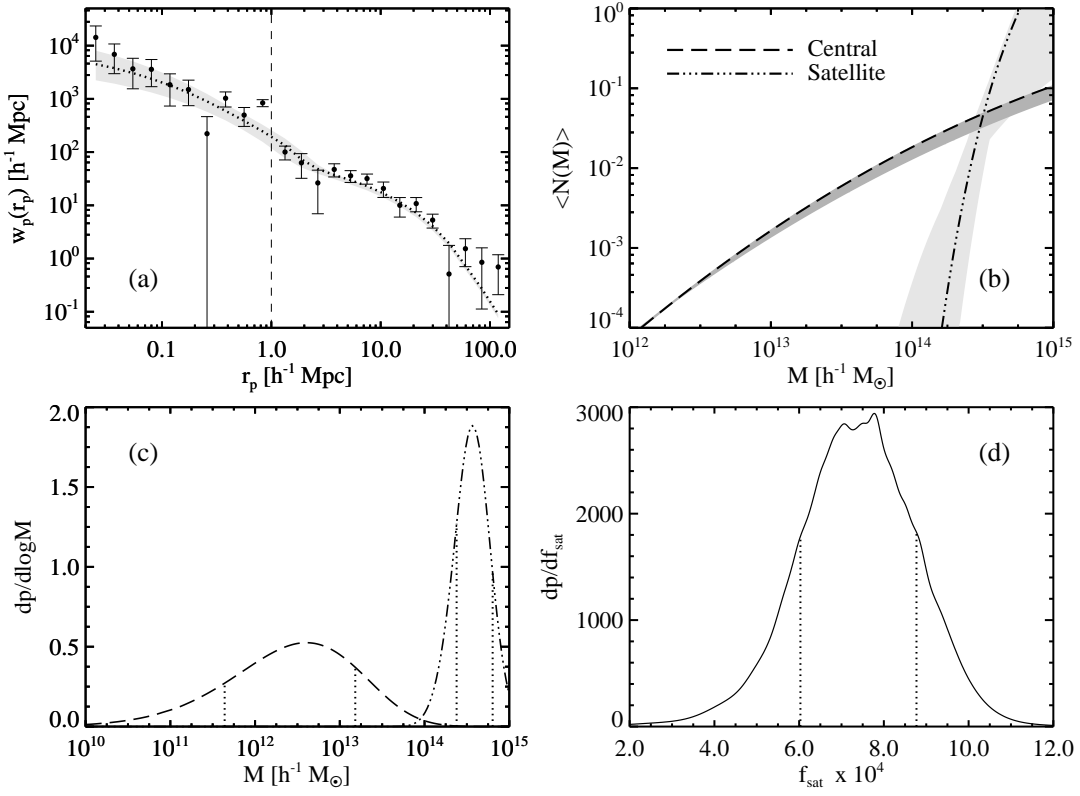}
\end{tabular}
\caption{
\label{fig:fig2}
Fit to the 2PCF of $\bar{z}=1.4$ quasars and the quasar HOD. Panel (a): our estimate of the full projected 2PCF of SDSS quasars (data points and error bars) against the prediction of our best-fit HOD model (dotted line). Panel (b): the mean occupation function of SDSS quasars, decomposed into its central (dashed line) and satellite (dot-dashed line) components. In both panel (a) and (b), the shaded envelopes indicate the $68\%$ confidence intervals (see the text). Panel (c): the probability distribution of host halo masses, shown for both central (dashed line) and satellite (dot-dashed line) quasars. These distributions are obtained by multiplying the mean occupation functions with the differential halo mass function, averaged over all the models in the MCMC chain (see $\S5.1$ for discussion). Panel (d): the probability density function of the satellite fraction as given by all our HOD models. In panels (c) and (d), the vertical dashed lines indicate the central $68\%$ for each distribution.
}
\end{center}
\end{figure*}

In Figure~\ref{fig:fig2} we show the HOD of SDSS quasars at $z \sim 1.4$. Panel (a) shows the projected 2PCF measurements of our low-redshift clustering sample against the theoretical prediction of our best-fit HOD model (dotted line). We identify the best-fit model as the point in our five-dimensional parameter space associated with the global $\chi^{2}$ minimum. If we rank the $\chi^2$ of all the models in ascending order, the first 68\% of models give a range of predicted $w_p$ indicated by the shaded envelope in Figure~\ref{fig:fig2}. We note that Poisson errors are quoted on data points below the SDSS fiber collision scale. In applying our correction for $\pi_{\mathrm{max}}$, the Poisson errors quoted by HE06 are adjusted to reflect the additional pair counts (see Table~2). Poisson counting errors assume that each quasar pair is statistically independent of all other pairs in the sample, which breaks down on large scales ($r_{p} \gtrsim 1 \; h^{-1} \; \mathrm{Mpc}$) because of correlations between pairs in different radial bins. This leads to an underestimate of the uncertainty (e.g., SH07). Hence we excluded the single HE06 data point at $r_{p} > 0.8 \; h^{-1} \; \mathrm{Mpc}$ from our modeling, which does not statistically alter the results. Our HOD model reproduces the clustering with a reduced $\chi^{2}$ of 1.05.

Panel (b) of Figure~\ref{fig:fig2} shows the mean occupation function from the best-fit HOD model, decomposed into its central (dashed line) and satellite (dot-dashed line) components. Similar to panel (a), the shaded regions indicate the range of the mean occupation function given by the $68\%$ of models with the smallest $\chi^2$. As seen in panel (b), the satellite occupation becomes significant at mass scales above $\sim 10^{14}$ $h^{-1}$ $\mathrm{M}_{\sun}$. This indicates that, at low redshift, typically only the most massive halos host multiple quasars. Panel (c) shows the full host halo mass distributions for central (dashed line) and satellite (dot-dashed line) quasars. It is derived by multiplying the mean occupation function of central (or satellite) quasars with the differential halo mass function. For a randomly chosen central (or satellite) quasar, the curve gives the probability distribution of its host halo mass. The curves are roughly log-normal for both central and satellite quasars. Finally, panel (d) shows the probability density function for the satellite fraction $f_{\mathrm{sat}}$ from all models in the MCMC chain. The satellite fraction is defined as the ratio of the number density of satellites (integrated over all halo masses) to the total number density of quasars (including both central and satellite quasars). At the $68\%$ confidence level, we find the satellite fraction of $z\sim 1.4$ SDSS quasars to be $f_{\mathrm{sat}}=(7.4 \pm 1.3) \times 10^{-4}$.

\subsection{Redshift Evolution}

\begin{figure}
\begin{center}
\begin{tabular}{c}.
\includegraphics[width=8cm]{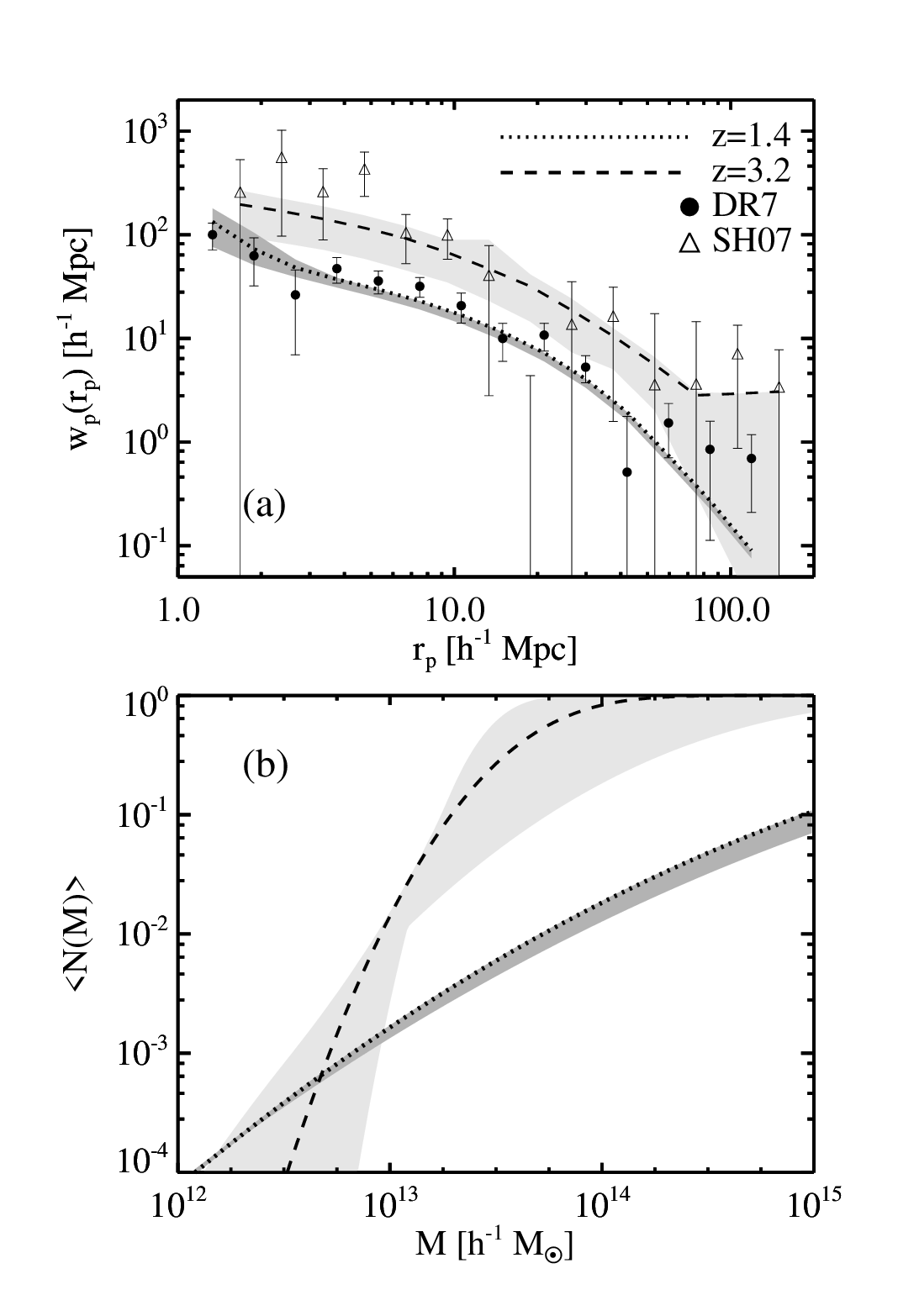} 
\end{tabular}
\caption{
\label{fig:fig3}
Redshift evolution of the central quasar HOD. Panel (a): projected 2PCF measurements and jackknife errors of SH07 (triangles) and those from SDSS DR7 (circles), together with the best-fit model prediction (dotted for $z\sim 1.4$ and dashed for $z\sim 3.2$). Panel (b): the best-fit mean occupation function of central quasars at $z\sim 1.4$ (dotted curve) and $z\sim 3.2$ (dashed curve). In both panels, the shaded region around 
each best-fit model shows the envelope determined from the 68\% of models with the smallest $\chi^2$ values.
}
\end{center}
\end{figure}

\begin{figure}
\begin{center}
\begin{tabular}{c}.
\includegraphics[width=8cm]{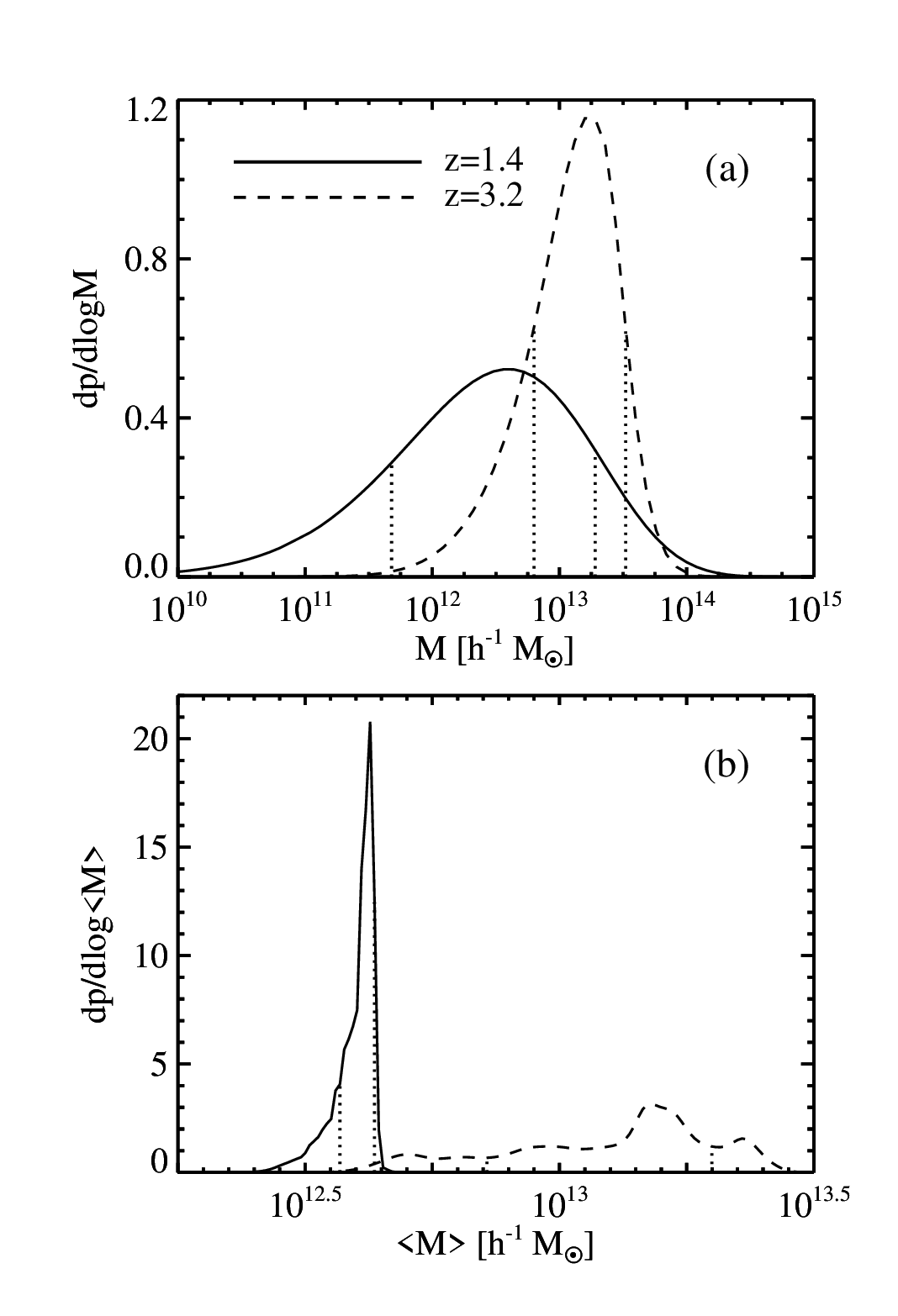} 
\end{tabular}
\caption{
\label{fig:fig4}
Redshift evolution of the host halo mass for central quasars. Panel (a): the full mass distribution of central quasar host halos, which is obtained by multiplying the central mean occupation function with the differential halo mass function at each redshift (see \S~5.1), averaged over all the models in the MCMC chain. Panel (b): probability density function of the median mass of halos hosting central quasars. The distribution is obtained by computing the median host halo mass for each model in the MCMC chain. In both panels, vertical lines denote the $68\%$ confidence intervals.
}
\end{center}
\end{figure}

By performing HOD modeling of the 2PCFs for both the $\bar{z}=1.4$ and $\bar{z}=3.2$ quasar samples, we are able to study the evolution of the quasar-halo relation.

Figure~\ref{fig:fig3} shows the redshift evolution of the SDSS central quasar population. In panel (a), triangles represent the projected 2PCF measurements of SH07 at high redshift and circles indicate our measurements from DR7 at low redshift. The best-fit theoretical predictions are represented by the dotted ($\bar{z}=1.4$) and dashed ($\bar{z}=3.2$) lines. As before, the shaded regions are the envelopes from predictions of the $68\%$ of models with the smallest values of $\chi^2$. Our best-fit model reproduces the high-redshift clustering with a reduced $\chi^{2}$ of 1.19. Panel (b) shows the best-fit mean occupation function of central quasars at $\bar{z} = 1.4$ (dotted line) and $\bar{z} = 3.2$ (dashed line). The mean occupation function steepens considerably with redshift over intermediate halo mass scales ($\sim 10^{13}-10^{14}$ $h^{-1}$ $\mathrm{M}_{\sun}$). The high-redshift occupation number exceeds the low-redshift one by a factor $> 10$ above masses of $\sim 10^{13.5}$ $h^{-1}$ $\mathrm{M}_{\sun}$. Since the cutoff profile reflects the scatter in the relation between halo mass and quasar luminosity, the above steepening implies that quasar luminosity is more tightly correlated with halo mass at higher redshift.

The mean occupation function shown in panel (b) of Figure~\ref{fig:fig3} can also be interpreted as the mass-dependent duty cycle of quasars (i.e., the fraction of halos with an active central quasar). At each redshift, we estimate an average duty cycle $f_{\mathrm{q}}$ for central quasars around the median host halo mass by averaging the mean central occupation number over the central $68\%$ of the host halo mass distribution. From all the models in the MCMC chains, we infer that, at the $68\%$ confidence level, $f_{\mathrm{q}}=7.3^{+0.6}_{-1.5}\times 10^{-4}$ for $z\sim 1.4$ quasars and $f_{\mathrm{q}}=8.6^{+20.4}_{-7.2}\times 10^{-2}$ for $z\sim 3.2$ quasars.

Panel (a) of Figure~\ref{fig:fig4} shows the normalized host halo mass distributions for central quasars at $\bar{z}=1.4$ (solid line) and $\bar{z}=3.2$ (dashed line). At high redshift we find both a narrowing of the distribution and a shift to higher masses. Panel (b) shows the probability density function for the median halo mass at each redshift. At the $68\%$ confidence level, denoted by the vertical lines under each curve, we find the characteristic host halo masses at $\bar{z} = 1.4$ and $\bar{z} = 3.2$ to be $M_{\mathrm{cen}} = 4.1^{+0.3}_{-0.4} \times 10^{12} \; h^{-1} \; \mathrm{M_{\sun}}$ and $M_{\mathrm{cen}} = 14.1^{+5.9}_{-6.9} \times 10^{12} \; h^{-1} \; \mathrm{M_{\sun}}$, respectively. There is tentative evidence for a lower quasar host halo mass at lower redshift, or the ``downsizing'' of quasar activity, but we note that this mass evolution is only significant at the $\sim1.5\sigma$ level. For satellite quasars, the characteristic median host halo mass at $\bar{z}=1.4$ is $M_{\mathrm{sat}} = 3.6^{+0.8}_{-1.0} \times 10^{14} \; h^{-1} \; \mathrm{M_{\sun}}$. The characteristic host halo mass for $\bar{z}=3.2$ quasars is not well constrained. We note that a luminosity difference exists between the high- and low-redshift samples. However, we find that the luminosity dependence of the clustering is not able to produce such a large shift in halo mass (see \S 6.2).

\section{Discussion}
We now discuss 
systematic issues and theoretical aspects of our analysis.

\subsection{Robustness of the Small-Scale Correlation Function}
At $r_p > 1 \; h^{-1}$ Mpc, quasars in both the HE06 sample and the SDSS DR7 sample are uniformly selected by the same targeting algorithm, indicating our correction for $\pi_{\mathrm{max}}$ should be complete. We have also verified that the redshift distribution of the corrected HE06 clustering sample is not appreciably altered from the original redshift distribution of HE06 sources. The consistency between the corrected HE06 measurements and the DR7 data then suggests that the effect on clustering from the remaining sample differences in redshift and luminosity does not exceed the $1\sigma$ level. 

At $r_p < 1 \, h^{-1}$ Mpc, we must examine the completeness of our correction for the larger $\pi_{\mathrm{max}}$, as the projected pair counts can only reflect the candidates identified by each selection algorithm for follow-up spectroscopy. Roughly $80\%$ of the sub-arcminute same-redshift quasar pairs were detected using the color-redshift relation (see HE06). Extending the line-of-sight separation limit from $\sim 22 \; h^{-1} \, \mathrm{Mpc}$ to $\sim 80 \; h^{-1} \, \mathrm{Mpc}$ corresponds to a redshift difference of only $\Delta z \sim 0.04$. For such a small redshift difference, the intrinsic dispersion and photometric redshift errors continue to dominate the scatter in the color-redshift relation, marginalizing the effect of our correction. This suggests that our correction with the SDSS DR7 $\pi_{\mathrm{max}}$ is reasonably complete. 

A second selection effect involves differences in luminosity threshold. The sub-arcminute clustering sample contains companion quasars primarily targeted to a fainter flux limit ($i < 21.0$) than that of the SDSS parent sample ($i < 19.1$). Although the luminosity evolution of small-scale quasar clustering has not yet been investigated, studies of large-scale clustering ($r_{p} > 1$ $h^{-1}$ Mpc) have detected weak to no luminosity evolution over similar redshift and luminosity ranges \citep[e.g.,][]{MY07a, SH09}. Assuming that the small-scale clustering evolves with luminosity similarly to large-scale clustering, we estimate that the amplitude of the projected 2PCF could be boosted, at most, by $\sim 30 \%$ for a flux limit change from $i < 21.0$ to $i < 19.1$. This is within the statistical uncertainty of our modeling. 

Finally, we must ensure that including the small-scale clustering (i.e., the corrected HE06 estimates) does not alter the constraints given by the large-scale DR7 measurements. The small-scale clustering mainly constrains the satellite HOD, while the central quasar HOD is mainly constrained by the large-scale clustering and the number density. We have verified that our modeling yields an identical central occupation function regardless of whether the small-scale data are included.

\subsection{Redshift and Luminosity Evolution of the HOD}

In order to have the strongest statistical power, our clustering samples are constructed over a broad redshift and luminosity range to maximize the volume and the number of sources. The 2PCF obtained in this way can be interpreted as an average over the redshift and luminosity intervals. However, interpreting the HOD as an average is problematic. Our HOD modeling uses halo properties (e.g., mass function, bias factor) and the quasar space density at the median redshift. Redshift evolution of the halo properties and quasar space density can lead to systematic effects that exceed the statistical uncertainties reflected in our modeling results. For a full interpretation of the measured 2PCF, all the evolution and selection effects should be properly incorporated into the model, which would require additonal HOD parameters. Given the above effects, it can seem that the interpretation of our HOD modeling result is not straightforward. However, we find that our modeling results can be meaningfully interpreted as representing the HOD for quasars at the median redshift of the sample (to within the quoted uncertainties), as explained and tested below.

The true 2PCF for quasars at the median redshift is that measured over a narrow redshift range around the median redshift. If the 2PCF measured over a wider redshift range around the median redshift (i.e., the average 2PCF) is statistically consistent with the true 2PCF for the median redshift, it would be reasonable to use the measured average 2PCF to represent the true 2PCF at the median redshift. The HOD modeling of such a 2PCF measurement using the halo properties and quasar abundance at the median redshift would then lead to HOD constraints that can be regarded as those for quasars at the median redshift.

To test whether our 2PCF measured over the full redshift range can be regarded as that at the median redshift, we divide the DR7 quasar sample into two subsamples above and below the median redshift and measure the projected 2PCF for each subsample. We find that the 2PCFs of the two sub-samples are consistent with each other and with the 2PCF of the full sample (i.e., the one we model). This indicates that the 2PCF measured from the full sample over the wide redshift range can indeed be regarded as the 2PCF of quasars at the median redshift, supporting our modeling of the measurements at this single redshift. Since we are not in a position to evaluate this assumption for the HE06 or SH07 samples, we assume the clustering evolves weakly on small scales and at high redshift, similarly to the large-scale clustering at low-redshift. This is not a strong assumption given the large relative errors on the corrected HE06 and SH07 measurements (typically $\gtrsim 50 \%$), which we expect to be sufficiently large to account for any biasing from redshift and/or luminosity evolution.

From our modeling, we find that the median mass of central quasar host halos at $\bar{z}=1.4$ is a factor of $\sim3.5$ lower than that at $\bar{z}=3.2$. Although the SDSS targets quasars at $z > 3.0$ to a fainter flux limit ($i < 20.2$) than those at $z < 3.0$ ($i < 19.1$), the high-redshift quasars in SH07 are still typically a factor of $\sim 2.8$ more luminous than those in our low-redshift DR7 sample. To check whether the luminosity difference can account for the shift in host halo mass scales, we perform the following test. If we adopt the luminosity dependent 2PCF results in \citet{SH09}, we find that increasing the median luminosity of the low-redshift quasars to match that of the high-redshift quasars can, at most, boost the amplitude of the low-redshift 2PCF by 50\%. HOD modeling of this boosted 2PCF indicates that the median mass of halos hosting central quasars is still a factor of 2.7 lower than that of the high-redshift sample. Therefore, we conclude that the difference in quasar luminosity cannot account for the halo mass scale evolution.

The decrease in this mass scale from $z\sim3.2$ to $z\sim1.4$ suggests some ``downsizing'' pattern in the evolution of the typical mass of quasar-hosting halos towards lower redshift, although the evidence is mild, significant only at the 1.5$\sigma$ level. If the luminosity dependence of clustering at high redshift is stronger than that in \citet{SH09}, the significance will be further reduced. Recently, some semi-analytic models and hydrodynamic simulations have shown that the luminosity evolution of quasar clustering is stronger at higher redshifts \citep[e.g.,][]{shen09, crotonetal09}. For such a case, the luminosity difference in our low- and high-redshift samples can result in a larger difference in host halo masses and the downsizing pattern may largely reflect a selection bias. Future measurements with larger quasar samples can better constrain the evolution of the quasar host halo mass.

\subsection{Alternate HOD Parameterizations}
Our parameterization of the quasar HOD is based on a physical model of AGN evolution from one hydrodynamic simulation \citep{Chatterjee12}. However, given the simulation volume, the model was limited to the case of low-luminosity AGN. Applying a similar parameterized form to quasars is a large extrapolation, so it is necessary to investigate whether our results depend on the particular HOD parameterization we have chosen and whether the HOD form is flexible enough to model the quasar 2PCF. 

In our fiducial HOD model, the central mean occupation function approaches unity at sufficiently large mass [see eq.~(\ref{eqn:2})], although in the modeling results it does not reach unity over the range of physical halo masses. To check whether our modeling results are affected by this asymptotic feature in our parameterization, we repeat our modeling and allow the softened step function to approach an arbitrary asymptotic value less than unity. With this additional degree of freedom, we find that the best-fit HOD is statistically identical to that of the original five-parameter model at both redshifts, indicating that our fiducial HOD parameterization has the flexibility similar to the six-parameter model in modeling the quasar 2PCFs here.

As a further test, we keep the above additional degree of freedom and modify the cutoff profile by changing the $1+{\rm erf}$ term in the central occupation function to $[1+{\rm erf}]^2$. Again, we are able to obtain good fits to the 2PCF measurements. Compared to the fiducial model, this modified model leads to a different amplitude of the mean central quasar occupation function for halos above $10^{13.5} \; h^{-1} \; M_\odot$. For $z\sim 1.4$ quasars, the amplitude from the best-fit model increases by a factor of $\sim 2$, while for $z\sim 3.2$ quasars, it drops by a factor of $\sim 15$. For the $z\sim 1.4$ quasars, the mass scale for satellite quasars shifts by about 0.5 dex towards the high mass end.

However, since massive halos are rare, the host halo mass distributions for central quasars remain statistically identical to those from the fiducial model (see Figure~\ref{fig:fig4}), indicating that our constraints on the typical host halo mass of central quasars are robust. The test suggests that the projected 2PCF has little constraining power for the central quasar HOD in high-mass halos. 

As an additional test of the lack of constraining power for the HOD in high mass halos, we introduce a sharp cutoff in the fiducial model $\langle N_{\mathrm{cen}}(M)\rangle$ for halos of mass $M>M_{\mathrm{crit}}$ 
and leave the form of $\langle N_{\mathrm{sat}}(M)\rangle$ unchanged. The motivation of such a high-mass cutoff is related to the decline in star formation for central galaxies in halos of $\gtrsim 10^{13} \; h^{-1} \; M_\odot$ \citep[e.g.,][]{conroy09} --- if both star formation and quasars are fueled by a common reservoir of cold gas, a decline in star formation would imply a decline in quasar activity in massive halos. For two tests with $M_{\mathrm{crit}}$=$10^{13} \; h^{-1} \; M_\odot$ and $10^{14} \; h^{-1} \; M_\odot$ we find equally good fits to the 2PCF (as compared to our fiducial modeling). The inferred satellite fractions and median host halo mass scales for central quasars are consistent with the results from the fiducial model within the uncertainty. The tests clearly show that the 2PCF is not sensitive to the central quasar HOD in
massive halos ($M\gtrsim 10^{13} \; h^{-1} \; M_\odot$). It also demonstrates that our results in central quasar halo mass scale are not strongly affected by the assumed form of the high-mass end of the HOD.

Unlike for central quasars, the satellite quasar HOD is affected more by the 
parameterization. The satellite fraction inferred from our fiducial model 
therefore has some systematic uncertainty, although those from the modified 
models remain consistent with that from the fiducial model at the $1\sigma$ 
level. It is certain, however, that at least a small fraction of quasars must 
be satellites to produce the small-scale clustering. Could all quasars be
satellites? Although this satellite-only scenario is not consistent with
observations of luminous AGN \citep{ST11}, we perform a test by setting 
$\langle N_{\mathrm{cen}}(M)\rangle=0$ and leaving the
rolloff power-law form of $\langle N_{\mathrm{sat}}(M)\rangle$ unchanged
so that $f_{\mathrm{sat}}=1$ \citep[e.g.,][]{MI11}. We find that the model is 
unable to simultaneously reproduce both the large- and small-scale clustering,
ruling out the possibility that all quasars are satellites (at least for the 
rolloff power-law form of the HOD adopted here).

Recently, after our initial submission of the paper, \citet{kayo12} 
reported a satellite fraction for $z\sim 1.4$ qusasars $\sim 100$ times higher 
than our inference, also based on HOD modeling of the 2PCF. They parameterize
the mean occupation functions of central and satellite quasars to have the
same shape and only differ in normalization. Both follow a Gaussian 
curve $\langle N(M)\rangle$ in logarithmic space of halo mass with fixed width 
($\sim 0.77$ dex in a full-width-half-maximum sense), and the relative 
normalization is given by the mass-independent satellite fraction 
$f_{\mathrm sat}$ such that 
$\langle N_{\mathrm{cen}}(M)\rangle=(1-f_{\mathrm{sat}}) \langle N(M)\rangle$
and 
$\langle N_{\mathrm{sat}}(M)\rangle=f_{\mathrm{sat}} \langle N(M)\rangle$. 
The two free parameters, $f_{\mathrm sat}$ and the halo mass scale 
are constrained by fitting the number density of quasars and the 2PCF.
In such a parameterization, $\langle N_{\mathrm{cen}}(M)\rangle$ is similar to our
test case with a high-mass cutoff, but $\langle N_{\mathrm{sat}}(M)\rangle$ 
is substantially different.

Although there is no compelling reason to believe that $\langle N_{\mathrm{cen}}(M)\rangle$ and $\langle N_{\mathrm{sat}}(M)\rangle$ have the same shape, the \citet{kayo12} model, differing substantially in 
$\langle N_{\mathrm{sat}}(M)\rangle$ from ours, can lead to good fits to the 
2PCF. This indicates that there exists large degeneracy in the satellite HOD 
constrained from the 2PCF. Therefore, compared to the central quasar
halo mass scale, the satellite HOD and satellite fraction are less robustly 
determined. Additional observables are necessary to break the degeneracy.
Since the satellite mass scales are different in our model and theirs, the
distribution of the line-of-sight velocity differences of quasar pairs can be 
used to differentiate the two models. \citet{kayo12} tabulate the relative 
line-of-sight velocities of quasar pairs in their sample. Although the current 
data are not precise enough, we make an attempt anyway to compare the velocity
distribution with model predictions. It appears that neither model is 
favored by the data. While our model produces too many pairs of large relative 
line-of-sight velocity (caused by putting too many satelltes in massive halos),
 theirs tends to underpredict such pairs. The true solution may lie in between 
the two models, and a larger sample of small-separation quasar pairs would help to definitively constrain the
satellite HOD. In conclusion, it is necessary to keep in mind the caveat of large degeneracy in interpreting the satellite HOD constraints in our and their modeling results. These degeneracies underscore the importance that future modeling employ physically-motivated parameterizations that allow the clearest possible interpretations.

Finally, we consider a parameterization in which the central quasar mean occupation function is modeled as a sharp step function with the amplitude as a free parameter. That is, this model assumes a constant duty cycle for central quasars. It also has five free parameters, but we find it is unable to reproduce the clustering on one-halo scales as a result of insufficient central-satellite pairs. As a caveat, we note that in our model we independently assign central and satellite quasars to halos (i.e., assume no correlation between the activities of the central and satellite black holes within a given halo; see \S 4.1). If major mergers trigger quasar activity, for example, there may be an enhanced probability of close quasar pairs in group-sized halos, suggesting satellite black holes are more likely to become active in a halo with an active central black hole. In such a case, preferentially assigning satellites to halos containing central quasars in principle may create sufficient central-satellite pairs to reproduce the one-halo clustering. However, we reserve a more detailed investigation for future work. Conservatively, we conclude that the constant duty cycle scenario for central quasars is excluded if the occupation statuses of central and satellite quasars are uncorrelated.

Overall, we find that our fiducial parameterization is able to well reproduce the observed quasar clustering and our modeling results on the mass scales of central quasars do not seem to be an artificial effect from the parameterization.

\section{Conclusions}

In this paper we present the first estimate of the projected 2PCF of SDSS quasars at $z \sim 1.4$ over the full one- and two-halo scales. We perform HOD modeling of the projected 2PCF to obtain tight constraints on the relation between central and satellite quasars and their host dark matter halos. We also model the projected 2PCF of SDSS quasars at high redshift ($z \sim 3.2$) measured by SH07 to investigate the redshift evolution of the HOD.

The key results from our study include:
\begin{itemize}

\item[(1)]
We measure the projected 2PCF of $z\sim 1.4$ SDSS quasars over a large range of scales, $0.02 \hinvMpc < r_p < 120 \hinvMpc$, by combining measurements from the SDSS DR7 data with those from the HE06 binary quasar 
sample (with appropriate corrections).

\item[(2)] 
Our quasar HOD model, with its parameterization motivated by AGN evolution simulation, is able to interpret the projected 2PCFs of SDSS quasars on all scales and at both $z\sim 1.4$ and $z\sim 3.2$.

\item[(3)] 
For the $z\sim 1.4$ SDSS quasar sample, we model the projected 2PCF over all scales to obtain the tightest constraints to date on the median host halo mass scales, $M_{\mathrm{cen}} = 4.1^{+0.3}_{-0.4} \times 10^{12} \; h^{-1} \; \mathrm{M_{\sun}}$ for central quasars and $M_{\mathrm{sat}} = 3.6^{+0.8}_{-1.0} \times 10^{14} \; h^{-1} \; \mathrm{M_{\sun}}$ for satellites. 

\item[(4)]
A small fraction of $z\sim 1.4$ SDSS quasars need to be satellites within dark matter halos to reproduce the small-scale clustering. Under our parameterization, we obtain the first estimate of the satellite fraction of $z\sim 1.4$ SDSS quasars, $f_{\rm sat}= (7.4 \pm 1.4) \times 10^{-4}$.

\item[(5)] 
The median host halo mass for $z\sim 3.2$ central quasars is inferred to be $M_{\mathrm{cen}} = 14.1^{+5.8}_{-6.9} \times 10^{12} \; h^{-1} \; \mathrm{M_{\sun}}$, about 3.5 times higher than that for $z\sim 1.4$ central quasars (a $1.5\sigma$ result).

\item[(6)]
The cutoff profile of the mean occupation function of central quasars steepens considerably with redshift. Since the cutoff profile reflects the scatter in the relation between halo mass and quasar luminosity, this steepening implies that quasar luminosity is more tightly correlated with halo mass at higher redshift.

\item[(7)] 
The average duty cycles for central quasars residing in halos around the median host halo mass are inferred to be $f_{\mathrm{q}}=7.3^{+0.6}_{-1.5}\times 10^{-4}$ at $z\sim 1.4$ and $f_{\mathrm{q}}=8.6^{+20.4}_{-7.2}\times 10^{-2}$ at $z\sim 3.2$.

\end{itemize}

The median mass scales inferred from our HOD analysis are consistent with the current paradigm of galaxy and quasar co-evolution \citep[e.g.,][]{hopkinsetal06, crotonetal09, hickoxetal09}. In this paradigm, a quasar phase and rapid growth of the stellar bulge of the galaxy occurs when the host halo reaches a critical mass between $10^{12}-10^{13}$M$_{\odot}$. Our results indicate a similar mass scale for both low- and high-redshift quasars, but more data is needed to precisely measure the redshift evolution of the host mass scales. A more accurate measurement of the mass scale will provide the necessary input for future developments and tests of quasar evolution theory.

Although the exact mechanism for triggering quasar activity is not fully known \citep[for a recent review, see][]{alexander11}, some of the leading candidates include higher gas inflow rates from gas-rich galaxy mergers \citep{k&h00,springeletal05b,hopkinsetal06}, secular processes and minor mergers \citep[e.g.,][]{k&k04}, and steady cold flows \citep{dimatteo12}. Better measurements of small-scale clustering at different redshifts may provide a useful way to understand triggering processes. For example, a purely merger-driven model will predict excess power at small scales compared to a model with only secular processes \citep[e.g.,][]{HO08, thackeretal09, degrafetal11a}. The halo mass scales inferred from HOD modeling can be used to predict the small-scale clustering of a population (e.g., galaxies) dominated by secular processes. With such a control sample, the relative small scale excess would provide important information on the role of mergers in triggering quasar activity. In a purely cold flow-driven model, quasar activity has a specific dependence on halo mass. Hence the mass scale constrained by modeling clustering measurements is also useful for testing such models.

We obtained a low satellite fraction for $z\sim 1.4$ quasars, while that for $z\sim 3.2$ quasars is not well constrained given the data. The redshift evolution of the satellite fraction, if constrained, can have important implications for black hole mergers and the binary AGN population. We find that the 2PCF alone may not be able to provide accurate and unbiased constraints on the satellite fraction of quasars, so other observables like 
the quasar pairwise velocity distribution would be useful for tightening the constraints.

The quasar duty cycle inferred from our study strongly varies with redshift similarly to predictions from semi-analytic models \citep[e.g.,][]{shen09}. A fair comparison between the exact values of duty cycles in different works is not straightforward, given the subtle differences in the definition and in the averaging. Broadly speaking, while we obtain a lower duty cycle at $z\sim1.4$, the duty cycle at $z\sim3.2$ is consistent with the predictions of \citet{Shankar10a}, who use a phenomenological model to describe the statistical properties of quasars. \citet{Chatterjee12} find a similar redshift evolution in the active fraction of low-luminosity AGN. Using a hydrodynamic simulation they show that feedback from the black hole suppresses its own growth, lowering the number density of AGN at low redshifts for comparable luminosities. Recently, \citet{Shankar10b} investigated the degeneracy in theoretical interpretation of the duty cycle with the quasar luminosity-host halo relation. Putting aside the subtlety in the exact definition, our low-redshift value of the duty cycle is consistent with \citet{Shankar10b}. In this model, a higher value of duty cycle corresponds to a lower scatter in the quasar luminosity-halo mass relation. Our results therefore support the picture of a lower scatter between halo mass and quasar luminosity at higher redshift.

Understanding the impact of AGN on cosmology and galaxy evolution is one of the key science drivers behind large-scale numerical simulations and future multi-wavelength surveys (e.g., BigBOSS, eROSITA, LSST, SDSS-III/BOSS). The HOD framework, capable of exploiting the statistical power of clustering in these simulations and survey data sets, is expected to provide increasingly informative tests of AGN evolution.

\section*{Acknowledgments}
We would like to thank Frank van den Bosch, Andrey Kravtsov, Adam Myers, Nikhil Padmanabhan, David Weinberg, and Martin White for useful discussions and comments. We thank Joe Hennawi for providing his data and for helpful comments. We also wish to thank the anonymous referee for useful suggestions that improved our paper. 
JR acknowledges support from a Yale College Dean's Science Research Fellowship through an early period of the project. ZZ gratefully acknowledges support from the Yale Center for Astronomy and Astrophysics through a YCAA fellowship at an early stage of this work. SC was partially supported by NSF grant AST-0806732 during the course of this work. DN was supported in part by NSF grant AST-1009811, NASA ATP grant NNX11AE07G, and by Yale University. YS acknowledges support from the Smithsonian Astrophysical Observatory (SAO) through a Clay Postdoctoral Fellowship.

\appendix

\section{Correcting the Small-Scale Correlation Function}
Because of the low pair counts at small separations, HE06 calculate their clustering measurements in terms of a dimensionless, volume- and redshift-averaged quantity, $\overline{W}_{p}$. They use the estimator
\begin{equation}
\label{eqn:A1}
1+\overline{W}_{p}(R_{\mathrm{min}},R_{\mathrm{max}})= \frac{ \langle QQ \rangle}{ \langle QR \rangle},
\end{equation}
where $\langle QQ \rangle$ is the data-data pair count assigned to the transverse radial bin $[R_{\mathrm{min}},R_{\mathrm{max}}]$ and $ \langle QR \rangle$ is the data-random pair count assigned to the same bin. The method of assigning $\langle QQ \rangle$ to each radial bin varies with targeting algorithm and $ \langle QR \rangle$ depends on a selection function defined to account for sources of incompleteness. The projected 2PCF, $w_{p}$, is recovered by integrating $\overline{W}_{p}$ along the line of sight,
\begin{eqnarray}w_{p} & = & \int_{-\frac{v_{\mathrm{max}}}{a(\bar{z}) H(\bar{z})}}^{\frac{v_{\mathrm{max}}}{a(\bar{z}) H(\bar{z})}} \overline{W}_{p}(R_{\mathrm{min}},R_{\mathrm{max}}) d\pi \nonumber \\& = & \frac{2v_{\mathrm{max}}}{a(\bar{z}) H(\bar{z})}\overline{W}_{p}(R_{\mathrm{min}},R_{\mathrm{max}}),
\end{eqnarray}
where we note $v_{\mathrm{max}}/[a(\bar{z}) H(\bar{z})] \sim \pi_{\mathrm{max}}$. HE06 set $v_{\mathrm{max}}$, the maximum velocity difference for a pair, to $2000 \; \mathrm{km  \; s}^{-1}$, which gives a rescaling factor of $2\pi_{\rm max} \sim 44\hinvMpc$.

\subsection{Correcting the Data-Data Pair Count}
To match the $\pi_{\rm max}$ used in the 2PCF measurement of the DR7 quasars, we correct the projected 2PCF measurements of HE06 by including the (previously excluded) projected quasar pairs with line-of-sight separations $< 80.0$ $h^{-1}$ Mpc, the upper line-of-sight separation limit of our DR7 measurements. HE06 publish a catalog of all their excluded projected pairs with a redshift difference of $\Delta z < 0.5$. Since $\pi_{\mathrm{max}}=80 \; h^{-1} \mathrm{Mpc}$ corresponds to $\Delta z \sim 0.04$, this sample is complete for our purposes. In applying the correction, we first select the projected pairs from these catalogs satisfying all the selection criteria of the particular targeting algorithm which identified them, excluding the velocity difference requirement $ \mid \Delta v \mid < 2000$ km $\mathrm{s}^{-1}$. We then calculate the line-of-sight separation for each selected pair of quasars from their spectroscopic redshifts. Each projected pair with $\pi < 80.0$ $h^{-1}$ Mpc is added to the appropriate transverse radial bin.

However, a subtlety arises in assigning the additional pair counts. For sub-arcminute pairs, the companion quasar is typically targeted to a fainter magnitude than its SDSS parent. Since auto-clustering measurements require a uniformly flux-limited sample, the usual interpretation of $\langle QQ \rangle$ and $\langle QR \rangle$ as pair counts cannot be applied. Instead, HE06 define $\langle QQ \rangle$ and $\langle QR \rangle$ in terms of the number of \textit{companions} about quasars in the parent samples. As we will show below, $\langle QR \rangle$ is then defined as dependent on the quasar luminosity function \citep[e.g.,][]{BO00,CR04a,RI05} to account for flux differences between the companion samples. For most quasar pairs selected at subarcminute separations, the parent and companion samples are distinct and the additoinal pair count is just the number of projected pairs we add to the bin. However, for overlapping pairs discovered in the parent SDSS + 2QZ catalog, the parent and companion samples are identical. In this case the additional pair count is \textit{twice} the number of projected pairs we add to the bin, since each of the two objects in the pair has a companion in the parent sample. We refer the reader to HE06 for a full description of this method.

\subsection{Correcting the Data-Random Pair Count}
We now turn our attention to the data-random pair count, $\langle QR \rangle$. Since companion quasars at $\theta \le 60''$ are targeted to magnitudes fainter than the SDSS flux limit, random mock catalogs cannot be used to estimate $\langle QR \rangle$ for each radial bin. Instead, HE06 separately compute the data-random contribution $\langle QR \rangle_{k}$ for each selection algorithm $k$ in terms of a selection function accounting for sources of incompleteness,
\begin{equation}
\langle QR \rangle_{k}= \sum_{j=1}^{N_{\mathrm{QSO}},\,k} n(z_{j}, \, i < i'_{k}) V_{\mathrm{shell}} S(z_{j}, \, \theta_{j}),
\end{equation}
where $n(z_{j}, \, i < i'_{k})$ is the number density of quasars above the flux limit $i'_{k}$ of selection algorithm $k$, $V_{\mathrm{shell}}$ is the cylindrical volume probed by the radial bin in redshift space, $S(z_{j}, \, \theta_{j})$ is the selection probability function, and $N_{QSO, \, k}$ is the total number of quasars in the parent sample of selection algorithm $k$. For each radial bin, the total $\langle QR \rangle$ is then obtained by summing over all contributions. In this calculation, only the cylindrical redshift-space volume probed by each radial bin, $V_{\mathrm{shell}} = 2 \pi (R_{\mathrm{max}}^{2}-R_{\mathrm{min}}^{2}) \pi_{\mathrm{max}}$, significantly depends on our choice of $\pi_{\mathrm{max}}$. Since, for each radial bin, $\langle QR \rangle \propto V_{\mathrm{shell}}$, our correction factor is given simply by the ratio of $\pi_{\mathrm{max}}$ values used in computing the projected 2PCF for our SDSS DR7 sample and the HE06 sample ($80 \; h^{-1} \; \mathrm{Mpc} / 22 \; h^{-1} \; \mathrm{Mpc} \approx 3.6$).

\subsection{Recalculating the Small-Scale Correlation Function}
With our corrected pair counts, we recalculate the small-scale projected 2PCF. First, we calculate the dimensionless clustering statistic $\overline{W}_{p}$ using the estimator of HE06 [see eq.~(\ref{eqn:A1})]. Then, we recover $w_{p}$ by integrating along the line of sight to $\pi_{\mathrm{max}} = 80 \; h^{-1} \; \mathrm{Mpc}$, the line-of-sight separation limit for our SDSS DR7 sample. Our complete correction to the projected 2PCF measurements of HE06 is given as
\begin{equation}
w_{p} = 2\pi_{\rm max} \left( \frac{ \langle QQ \rangle_c}{ \langle QR \rangle_c} - 1 \right),
\end{equation}
where $\langle QQ \rangle_c$ is the corrected data-data pair count for each radial bin as described above and $\langle QR \rangle_c = (\pi_{\rm max}/ \pi_{\rm max,0}) \langle QR \rangle $ is the corrected data-random pair count, with $\pi_{\rm max}=80 \; h^{-1} \; \mathrm{Mpc}$ and $\pi_{\rm max,0} = 22 \; h^{-1} \; \mathrm{Mpc}$. Finally, we re-estimate the Poisson errors using the corrected data-data counts. Following HE06, we use the uncertainties tabulated in \citet{GE86} for bins with $\langle QQ \rangle_c < 30$.

\bibliography{hodbib}{}
\bibliographystyle{apj}

\end{document}